\documentclass[12pt]{article}
\usepackage{wider,scicite}
\begin{document}

\newcommand{\sqvb}{\ensuremath{ \langle \!\langle 0 |} }
\newcommand{\sqvk}{\ensuremath{ | 0 \rangle \!\rangle } }
\newcommand{\sqvn}{\ensuremath{ \langle \! \langle 0 |  0 \rangle \! \rangle} }
\newcommand{\wh}{\ensuremath{\widehat}}
\newcommand{\be}{\begin{equation}}
\newcommand{\ee}{\end{equation}}
\newcommand{\bea}{\begin{eqnarray}}
\newcommand{\eea}{\end{eqnarray}}
\newcommand{\ra}{\ensuremath{\rangle}}
\newcommand{\la}{\ensuremath{\langle}}
\newcommand{\rra}{\ensuremath{ \rangle \! \rangle }}
\newcommand{\lla}{\ensuremath{ \langle \! \langle }}
\newcommand{\str}{\rule[-.125cm]{0cm}{.5cm}}
\newcommand{\pr}{\ensuremath{^\prime}}
\newcommand{\ppr}{\ensuremath{^{\prime \prime}}}
\newcommand{\da}{\ensuremath{^\dag}}
\newcommand{\as}{^\ast}
\newcommand{\eps}{\ensuremath{\epsilon}}
\newcommand{\ve}{\ensuremath{\vec}}
\newcommand{\ka}{\kappa}
\newcommand{\non}{\ensuremath{\nonumber}}
\newcommand{\lf}{\ensuremath{\left}}
\newcommand{\rt}{\ensuremath{\right}}
\newcommand{\al}{\ensuremath{\alpha}}
\newcommand{\dfn}{\ensuremath{\equiv}}

%citesupernumberparens.tex: Parentheses around superscript added by Mark A. Rubin 7/11/02
%rubin@ll.mit.edu
%
% version = 1.02 of citesupernumber.sty 1998 May 3
% origin 1989 January 20
% originally named citsupernumber.tex, but citesupernumber.sty works
%
% call in LaTeX2e:
% \usepackage{citesupernumber}
%
% This file makes citations be superscripted.  They come out as numbers.
% For example, by putting the following code in your LaTeX file,
%
%     text text text\cite{Darwin1859,Clark2001} text ...
%     \input citesupernumber
%     moretext moretext moretext\cite{Darwin1859,Clark2001} moretext ...
%
% you will get:
% 
%     text text text[1,2] text ...
%
%                              1,2
%     moretext moretext moretext   moretext ...

% NOTE:  By making this a package, the above switch does not work!!
% But then, would one really want to switch like that?

%
% This code was originally made by 
% Michael DeCorte // (315)265-2439 // P.O. Box 652, Potsdam, NY 13676
% Internet: mrd@sun.soe.clarkson.edu  // Bitnet:   mrd@clutx.bitnet        
% 
% It was modified by Tom Schneider toms@ncifcrf.gov

% DeCorte's original code was:
% \def\@cite#1#2{$^{#1\if@tempswa , #2\fi}$}
% I have added the \scriptsize so that the numbers are appropriately smaller
% and the mbox to make scriptsize work under the mathmode that is
% used to make the superscripting.
%

\makeatletter
\def\@cite#1#2{$^{\mbox{\scriptsize (#1\if@tempswa , #2\fi)}}$}
\makeatother

% ---------------------------------------------------
% If the following line:
% \def\cite#1{$^{#1}$}
% is also used, the output would be:
%                               Darwin1859,Clark2001
%     moretext moretext moretext                    moretext ...
%
% ---------------------------------------------------
% The next improvement of this would be to make it more like a
% switch, so one would say \citesupernumber or \citesuperkey
% or \citebracket or \citeparenthesis to get different forms.
% Better yet would be \citationform{place}{kind}{edge}
% where
%       place is 'inline' or 'superscript'
%       kind is 'citationkey' or 'number'
%       edge is 'bracket' or 'parenthesis' or 'none'
% Do you know how to make this happen?
% 
% While we are into wishing, it would be useful for some journals
% to have the following citation forms built, not from the key,
% but from the year and author list:
% One author:   (Smith, 1982)
% Two authors:  (Smith and Jones 1892)
% More than two authors:  (Smith {\em et al}, 1982)
% with ';' as separators between the citations.
% (For use in Journal of Molecular Biology.)

\title{\bf Locality  in the Everett Interpretation of 
Quantum Field Theory\thanks{
This work was sponsored by the Air Force under Air Force Contract 
F19628-00-C-0002.  Opinions, interpretations, conclusions, and 
recommendations are those of the author and are not necessarily endorsed 
by the U.S. Government.
}
}
\author{
Mark A. Rubin\\
\mbox{}\\   %REMOVE FOR DOUBLE-SPACE  
Lincoln Laboratory\\ 
Massachusetts Institute of Technology\\  
244 Wood Street\\                         
Lexington, Massachusetts 02420-9185\\      
rubin@LL.mit.edu\\ 
}
\date{\mbox{}}
\maketitle

\begin{abstract}
%\begin{quote}
%\small
%
%\mbox{}
%
%%\hspace*{.25in}
%\hspace*{.15 in}
Recently it has been shown that transformations  of Heisenberg-picture
operators are the causal mechanism which allows Bell-theorem-violating
correlations at a distance to coexist with locality in the Everett interpretation
of quantum mechanics.   A calculation to first order in perturbation theory
of the generation of EPRB entanglement in nonrelativistic fermionic field theory 
in the Heisenberg picture illustrates that the same mechanism leads to correlations 
without nonlocality in quantum field theory as well. An explicit transformation 
is given to a representation in which initial-condition information
is transferred from the state vector to the field operators, making the
locality of the  theory  manifest.\\

\end{abstract}

\section{INTRODUCTION}

Bell's theorem\cite{Bell64} does not apply to quantum mechanics in the Everett 
interpretation.\cite{Everett57}
The premises of the theorem include the implicit 
assumption that, each time an experiment is performed, there is a single outcome 
to the experiment.  In the Everett interpretation, all possible outcomes occur.
Thus, Everett-interpretation quantum mechanics is not demonstrated by Bell's theorem
to be nonlocal.\cite{Stapp80,Stapp85}\footnote{Locality
in Everett-type interpretations of quantum mechanics is also  discussed in
Refs. 2, 5-17.}
%\cite{Everett57}, \cite{Zeh70}-\cite{Vaidman01}.}

It is another question, however, whether Everett-interpretation quantum mechanics {\em is}\/
in fact local---that is, whether or not it contains a local
realistic mechanism for producing correlations observed between distant systems which have interacted in the past, such as those
observed  in Einstein-Podolsky-Rosen-Bohm (EPRB) 
experiments.\cite{EPR,Bohm51}  In other interpretations of
quantum mechanics the view is taken that each repetition of an experiment 
 results in a single outcome. In the context of these other 
interpretations we attempt to understand how it comes about
that when the experimenters, Alice and Bob, set their respective analyzer magnets
to be parallel, the results which they obtain for the respective spins of a
pair of singlet-state electrons are perfectly anticorrelated.  Bell's theorem
states that neither quantum mechanics nor any other conceivable physical
theory can provide a causal explanation for them and, at the same time,
account for the correlations observed when the magnets are not parallel.

In the Everett interpretation, correlations between the two experimenters' results are not at issue; rather, a different question of causation arises. 
According to Everett, both possible outcomes, spin-up and spin-down,
occur at each analyzer magnet and, at the conclusion of the experiment, there
are  two copies of each experimenter.\footnote{Or, in another variant of the
Everett interpretation, two continuously-infinite sets of copies of each 
experimenter.\cite{Deutsch85}} When they
 compare their respective results using some causal
means of communication, 
Alice-who-saw-spin-up only talks to   Bob-who-saw-spin-down,
and Alice-who-saw-spin-down always converses  with Bob-who-saw-spin-up.  What is the mechanism which brings about this perfect anticorrelation
in the possibilities for exchange of information between the Alices and the Bobs?

Deutsch and Hayden\cite{DeutschHayden00} have identified this mechanism.
In the Heisenberg picture of quantum mechanics, the properties of physical systems
are represented by time-dependent operators. When two systems interact, the operators
corresponding to the properties of  each of the systems  may acquire nontrivial tensor-product
factors acting in the state space of the other system.  These factors are in effect labels, 
appending to each system a record of the fact that it has interacted with a certain other 
system in a certain way.\cite{Rubin01} So, for example, when the two  
particles in the EPRB experiment are initially prepared in the singlet state, the interaction
involved in the preparation process causes the spin operators
of each particle to  contain nontrivial factors acting in the space in which the spin
operators of the other particle act. When Alice measures the spin of one of the
particles, the operator representing her state of awareness ends up with factors
which act in the state space of the particle which she has measured, as well as
in the state space of the other particle.  The operator corresponding to Bob's
state of awareness is similarly modified.  When the Alices and Bobs meet to compare
notes, it is these factors which lead to the correct pairing-up of the four of
them.

The amount of information which even a simple electron carries with it regarding 
the other particles with which it has 
interacted is thus enormous.
In Ref. 22  %\cite{Rubin01} 
I termed this the problem of ``label proliferation,'' and suggested that the
physical question of how all this information is stored\footnote{There is no {\em formal}\/ problem in this regard.  The mathematical elements which  encode the label information, the
tensor-product factors, are a straightforward  consequence of the basic
rules of quantum mechanics in the Heisenberg picture.} might receive an 
answer in the framework of quantum field theory. More generally, quantum field theory
is a  description of nature encompassing a wider range of physical phenomena
than the quantum mechanics of particles;  it is therefore of interest to investigate
the degree to which the conceptual picture of the labeling mechanism for bringing about
correlations at a distance in a causal manner accords with the field-theoretic 
formalism.

Indeed, there is a simple line of argument
which leads to the conclusion that 
Everett-interpretation Heisenberg-picture quantum field 
theory {\em must}\/ be local.  The dynamical
variables of the theory are field operators 
defined at each point in space, whose dynamical
evolution is described by local (Lorentz-invariant,
 in the relativistic case) differential
equations.  And the Everett interpretation
removes nonlocal reduction of the wavefunction
from the formalism. So how can nonlocality enter 
the scene?  

   This argument as it stands is 
incorrect, but it can be modified so that its 
conclusion, the locality of Everett-interpretation 
Heisenberg-picture quantum
field theory, holds.  What is wrong is
the following: While it is certainly true that 
operators in Heisenberg-picture quantum field 
theory evolve according to local differential 
equations, it is {\em not}\/ in general true that
all of the information needed to determine the 
outcomes and probabilities of measurements is 
contained in these operators.  Initial-condition
information, needed to determine probabilities,
resides in the time-independent Heisenberg-picture 
state vector. Since not all information is 
carried in the operators, is in incorrect to
argue that the local evolution of the operators
implies locality of the theory.

    However, as discussed in Sec. 4 below, 
it turns out to be possible
to transform from the usual representation of
the Heisenberg-picture field theory to other
representations in which the operators also 
carry the initial-condition information.  
So, in these representations,
the simple argument above for the locality of
Heisenberg-picture quantum field theory
is valid.  Bear in mind that in
these representations the use of the Everett 
interpretation still is crucial for the theory to
be local. As mentioned above, the Everett 
interpretation removes a source of explicit 
nonlocality in the theory (wavefunction collapse);
it ``defangs'' the Bell argument that, 
notwithstanding the explicitly local transfer
of information in the operators, {\em something}\/
else of a nonlocal nature must be going on;
and it provides labeling as
an alternative to the ``instruction set''
mechanism which in single-outcome
interpretations appears as the only explanation 
for correlations-at-a-distance  and which is what 
ultimately leads to Bell's theorem. This last 
issue of instruction sets and 
labels is no different in field theory than in 
first-quantized theory, and is discussed in Ref 22.
%\cite{Rubin01}. 
In field theory as in 
first-quantized theory, interaction-induced 
transformations of Heisenberg-picture 
operators (field operators, of course,
in the field theory case---see e.g., eq. 
(\ref{Vpertsol}) below) serve to encode the label 
information.

For simplicity, we will consider below only nonrelativistic quantum fields of
spin 1/2 fermions, and we will not explicitly include in the formalism
degrees of freedom corresponding  to the states of awareness of observers.
In Sec. 2 below I recast the first-quantized
analysis of the EPRB experiment presented in Ref. 22
%\cite{Rubin01}  
into a Fock-space formalism
of spinning particles with no spatial degrees of freedom. This analysis is generalized  in
Sec. 3 to nonrelativistic quantum field theory in three space dimensions. 
A perturbative calculation of the generation 
of EPRB entanglement provides an illustrative 
example of the absence of initial-condition
information from the usual Heisenberg-picture
field operators, as well as an explicit
formula for the degree of entanglement
for narrow wavepackets of massive particles.
In Sec. 4 a unitary transformation is presented which transfers initial-condition
information from the Heisenberg-picture state vector to the operators. 
In Sec. 5 I
review the conclusions that follow from this analysis  regarding the questions of label proliferation and locality.

\section{EPRB ENTANGLEMENT IN FOCK SPACE}

To construct a Fock space representation of the particles involved
in the EPRB experiment, we will first set up a suitable first-quantized
Schr\"{o}dinger-picture formalism, then proceed to the corresponding Fock space
representation by standard methods (see, e.g., Refs. 23-25).%\cite{Dirac}-\cite{Peres93}).

\subsection{First-Quantized Formalism}

\subsubsection{Single-Particle Hilbert Space}

We first consider a Hilbert space for a single spin 1/2 particle which 
has no spatial degrees of freedom but which does posses one
additional internal two-valued degree of freedom, which we will refer to
as its ``species.''  If the operator $\wh{\rho}$\/ corresponds to a determination
of the species of the particle, and $\wh{\sigma}_z$\/ measures the $z$\/-component
of spin in units of $\hbar/2$\/, a complete set of eigenstates for this system, ${\cal S}$\/,
is 
\be
|{\cal S}; \rho_{[r]},\alpha_i \ra , \hspace*{5mm} r,i=1,2,
\ee
where
\be
|{\cal S}; \rho_{[r]},\alpha_i \ra \dfn |{\cal S}; \rho_{[r]} \ra 
                                     |{\cal S}; \alpha_i \ra , \label{statedef}
\ee
\be
\wh{\rho} |{\cal S};\rho_{[r]} \ra = \rho_{[r]}|{\cal S};\rho_{[r]}\ra, 
\ee
\be
\wh{\sigma}_z |{\cal S}; \alpha_i \ra = \alpha_{i}|{\cal S};\alpha_i \ra, \label{sigmazaction}
\ee
\be
\alpha_1=1, \ \alpha_2=-1. \label{alphavalues}
\ee
Square brackets around subscripts denote indices of species eigenvalues.
The species eigenvalues are nondegenerate but otherwise arbitrary:
\be
\rho_{[1]} \neq \rho_{[2]}. \label{nondegenerate}
\ee
 For later
reference we note here the action of the spin operators
in the $x$\/ and $y$\/ directions,  $\wh{\sigma}_x$\/ and $\wh{\sigma}_y$\/,
on $|{\cal S}; \alpha_i \ra$\/:
\bea
\wh{\sigma}_x |{\cal S}; \alpha_i \ra &=& |{\cal S};\alpha_{\bar{i}} \ra, 
\label{sigmaxaction} \\
\wh{\sigma}_y |{\cal S}; \alpha_i \ra &=& i\alpha_{i}|{\cal S};\alpha_{\bar{i}}\ra,
\label{sigmayaction} 
\eea
where $\bar{i}$\/ denotes the complement of $i$\/,
\be
\bar{1}=2, \ \bar{2}=1. \label{complement}
\ee

\subsubsection{Two-Particle Hilbert Space}

The Hilbert space for two distinguishable particles is the  tensor product of 
two single-particle Hilbert spaces, spanned by the eigenvectors
\be
|{\cal S}^{(1)}{\cal S}^{(2)}; \rho_{[r]}, \alpha_{i}, \rho_{[s]}, \alpha_{j} \ra 
\equiv |{\cal S}^{(1)}; \rho_{[r]} ,\alpha_{i} \ra  |{\cal S}^{(2)}; \rho_{[s]}, \alpha_{j} \ra \hspace*{5mm} r,i,s,j=1,2,
\ee
where
\be
\wh{\rho}^{(1)} |{\cal S}^{(1)}{\cal S}^{(2)}; \rho_{[r]} ,\alpha_{i}, \rho_{[s]}, \alpha_{j} \ra  = \rho_{[r]}|{\cal S}^{(1)}{\cal S}^{(2)}; \rho_{[r]}, \alpha_{i} ,\rho_{[s]} ,\alpha_{j} \ra , 
\ee
\be
\wh{\rho}^{(2)} |{\cal S}^{(1)}{\cal S}^{(2)}; \rho_{[r]} ,\alpha_{i}, \rho_{[s]}, \alpha_{j} \ra  = \rho_{[s]}|{\cal S}^{(1)}{\cal S}^{(2)}; \rho_{[r]}, \alpha_{i}, \rho_{[s]} ,\alpha_{j} \ra , 
\ee
\be
\wh{\sigma}_z^{(1)} |{\cal S}^{(1)}{\cal S}^{(2)}; \rho_{[r]}, \alpha_{i}, \rho_{[s]} ,\alpha_{j} \ra  = \alpha_{i}|{\cal S}^{(1)}{\cal S}^{(2)}; \rho_{[r]},\alpha_i ,\rho_{[s]} ,\alpha_{j} \ra ,
\ee
\be
\wh{\sigma}_z^{(2)} |{\cal S}^{(1)}{\cal S}^{(2)}; \rho_{[r]}, \alpha_{i}, \rho_{[s]} ,\alpha_{j} \ra  = \alpha_{j}|{\cal S}^{(1)}{\cal S}^{(2)}; \rho_{[r]},\alpha_i, \rho_{[s]}, \alpha_{j} \ra .
\label{sigmaz2action}
\ee
Superscript indices specifying particles are in parentheses.

\subsubsection{Physical States}

The physical two-particle states are the  states in the two-particle
Hilbert space above which are antisymmetric under interchange of particle index.
This subspace is spanned by the (overcomplete) set of states 
\be
|[r],i,[s],j \ra  \equiv \frac{1}{\sqrt{2}} \left( |{\cal S}^{(1)}{\cal S}^{(2)}; \rho_{[r]}, \alpha_{i}, \rho_{[s]}, \alpha_{j} \ra - |{\cal S}^{(1)}{\cal S}^{(2)}; \rho_{[s]}, \alpha_{j}, \rho_{[r]},\alpha_{i} \ra \right), \hspace*{5mm} r,i,s,j=1,2. \label{spin_species}
\ee
We will  be particularly interested in  the four-dimensional subspace of physical states 
with one particle of each species.\footnote{Had we restricted ourselves to a single species,
we would have at our disposal at this point only a single physical state, 
$(1/\sqrt{2})\lf(|{\cal S}^{(1)}; \alpha_1\ra
                       |{\cal S}^{(2)}; \alpha_2\ra
                      -|{\cal S}^{(1)}; \alpha_2\ra
                       |{\cal S}^{(2)}; \alpha_1\ra\rt)$\/.
}
We will use for this subspace the basis
\be
|[1],i,[2],j\ra, \hspace*{5mm}  i,j=1,2. \label{spin_spec_corr_states}
\ee
The operator which corresponds to a measurement of the total spin of
the particles of species $r$\/ is
\be\
\wh{\vec{\sigma}}_{[r]} = |{\cal S}^{(1)} ; \rho_{[r]} \ra 
\la {\cal S}^{(1)} ; \rho_{[r]} | \otimes \wh{\vec{\sigma}}^{(1)} 
\otimes \wh{I}^{(2)}
+ \wh{I}^{(1)}\otimes |{\cal S}^{(2)} ; \rho_{[r]} \ra 
\la {\cal S}^{(2)} ; \rho_{[r]} |\otimes \wh{\vec{\sigma}}^{(2)} \label{sigmasubr}
\ee
The action of $\wh{\sigma}_{[r],z}$\/ on the states (\ref{spin_spec_corr_states}) is
\be
\wh{\sigma}_{[r],z}|[1],i,[2],j\ra
=\lf(\delta_{r1}\al_i + \delta_{r2}\al_j\rt)|[1],i,[2],j\ra.
\ee
So, spin and species are correlated in these states.
States $|J,J_z\ra $ \/ of definite (spin) angular momentum can be constructed
out of them:
\begin{eqnarray}
|1,1\ra &  =  & |[1],1,[2],1\ra , \label{initial_state}\\
|1,0\ra & =  & \frac{1}{\sqrt{2}}\left ( \str  |[1],1,[2],2\ra  + |[1],2,[2],1\ra \right), \\
|0,0\ra & =  & \frac{1}{\sqrt{2}}\left ( \str  |[1],1,[2],2\ra  - |[1],2,[2],1\ra \right), \label{singlet_state} \\
|1,-1\ra &  =  & |[1],2,[2],2\ra ,
\end{eqnarray}
where
\begin{eqnarray}
\wh{\vec{j}} \cdot \wh{\vec{j}}|J,J_z\ra & = & J(J+1)|J,J_z\ra, \\
\wh{j}_z|J,J_z\ra & = & J_z |J,J_z\ra, 
\eea
 $\wh{\vec{j}}$\/ being the total angular momentum operator,
\be
\wh{\vec{j}}  =  \left ( \wh{\vec{\sigma}}^{(1)} + \wh{\vec{\sigma}}^{(2)} \right)/2.
\ee

\subsubsection{Time Evolution and Entanglement}

In Ref. 22 
%\cite{Rubin01}  
a system of two distinguishable particles, each with
spin 1/2 and with no other dynamical properties, is 
considered. That is, the Hilbert space of the $p^{th}$\/ particle is spanned by
the two eigenvectors of the $z$\/ component of its spin, 
$|{\cal S}^{(p)}; \alpha_1\ra $\/ and   $|{\cal S}^{(p)}; \alpha_2\ra $\/.

These particles are prepared in the initial unentangled state
\be
|{\cal S}^{(1)};\alpha_1\rangle |{\cal S}^{(2)};\alpha_2\rangle, \label{initial_state_dist}
\ee
where particle 1 has spin up and particle 2 has spin down.
The particles are subsequently subjected to the action of the unitary time evolution operator 
$ \widehat{u}_{E-dist} $\/ which, in the Schr\"{o}dinger picture, has the effect
 of taking the state (\ref{initial_state_dist})
to the entangled singlet state ($J=J_z=0$\/):
\be
\widehat{u}_{E-dist} |{\cal S}^{(1)};\alpha_1\rangle |{\cal S}^{(2)};\alpha_2\rangle 
= \frac{1}{\sqrt{2}}\left( 
|{\cal S}^{(1)};\alpha_1\rangle |{\cal S}^{(2)};\alpha_2\rangle - 
|{\cal S}^{(1)};\alpha_2\rangle |{\cal S}^{(2)};\alpha_1\rangle
\right) \label{singlet_state_dist}
\ee

The state (\ref{initial_state_dist}) is considered unentangled because it is the
product of a vector in the state space of particle 1 and a vector in the state 
space of particle 2.  This notion of entanglement is inapplicable to
the  fermions with which we are concerned here, since any physical state of
such particles must be antisymmetric under the interchange of the indices labeling
the particles.\cite{Peres93} However, while it is not possible 
to distinguish particle 1 from
particle 2, it it possbile to distinguish a particle of species 1 from a particle
of species 2.
We will therefore take the initial
unentangled state of the two particles to be $|[1],1,[2],2\ra$\/, containing a particle
of species 1 with spin up and  particle of species 2 with spin down. The time-evolution
operator, $\wh{u}_E$\/, will be such as to transform $|[1],1,[2],2\ra$\/ to the singlet
state (\ref{singlet_state}); specifically,
\begin{eqnarray}
\wh{u}_E |[1],2,[2],1\ra & = &  |1,0\ra \label{uEaction}\\
\wh{u}_E |[1],1,[2],2\ra & = &  |0,0\ra \\
\wh{u}_E |[r],i,[s],j\ra & = &  |[r],i,[s],j\ra, \non \\ %\hspace*{5mm} 
&&\la[1],2,[2],1|[r],i,[s],j\ra =\la[1],1,[2],2|[r],i,[s],j\ra =0.
\end{eqnarray}
This operator can be written as
\be
\wh{u}_E=\wh{u}_E(\pi /4),
\ee
where
\begin{eqnarray}
\wh{u}_E(\gamma) & = &\exp(-i\gamma \wh{g}) \nonumber\\
                 & = &\wh{I}^\prime +\cos(\gamma) \wh{I}_2 
                                    -i \sin(\gamma) \wh{g}. \label{u_E_1Q_action}
\label{uEdef}
\end{eqnarray}
Here $\wh{I}_2$\/ is the projection operator into the two-dimensional
subspace upon which $\wh{u}_E$\/ acts nontrivially,
\be
\wh{I}_2= |[1],2,[2],1 \ra \la [1],2,[2],1| + |[1],1,[2],2 \ra \la [1],1,[2],2|,
\ee
$\wh{I}^\prime $\/ is the projection operator into the rest of the state
space,
\be
\wh{I}^\prime=\wh{I}-\wh{I}_2,
\ee
($\wh{I}=$\/ identity operator), and the generator of entanglement
$\wh{g}$\/ is given by
\be
\wh{g}=i\left(\str |[1],1,[2],2\ra  \la [1],2,[2],1 | -
                   |[1],2,[2],1\ra  \la [1],1,[2],2 |\right). \label{gopdef}
\ee

\subsubsection{Spin Correlation}

The operator which
corresponds to a measurement of the product of the projection of the spin of the species 1 particle
along the unit vector $\vec{n}_1$\/ with the projection of the spin of the
species 2 particle along the unit vector $\vec{n}_2$\/, is, using (\ref{sigmasubr}),
\bea
\wh{\xi} &=& ( \vec{n}_1 \cdot \wh{\vec{\sigma}}_{[1]})( \vec{n}_2 \cdot \wh{\vec{\sigma}}_{[2]}) \non \\
          &=&  |{\cal S}^{(1)} ; \rho_{[1]} \ra 
\la {\cal S}^{(1)} ; \rho_{[1]} | \otimes \vec{n}_1 \cdot \wh{\vec{\sigma}}^{(1)} 
\otimes \ |{\cal S}^{(2)} ; \rho_{[2]} \ra 
\la {\cal S}^{(2)} ; \rho_{[2]} | \otimes \vec{n}_2 \cdot \wh{\vec{\sigma}}^{(2)} \non \\
 & & +  |{\cal S}^{(1)} ; \rho_{[2]} \ra 
\la {\cal S}^{(1)} ; \rho_{[2]} | \otimes \vec{n}_2 \cdot \wh{\vec{\sigma}}^{(1)}
\otimes \ |{\cal S}^{(2)} ; \rho_{[1]} \ra 
\la {\cal S}^{(2)} ; \rho_{[1]} | \otimes \vec{n}_1 \cdot \wh{\vec{\sigma}}^{(2)} .
\label{spincorr} 
\eea

The spin correlation at the initial time $t_0$\/, when the particles are in
the  state $|[1],1,[2],2\ra$\/, is 
\be
C_{1Q}(0)  = \la  [1],1,[2],2 |\,\wh{\xi}\,|[1],1,[2],2 \ra . 
\ee
Using (\ref{statedef})-(\ref{spin_spec_corr_states}) and (\ref{spincorr}),
\be
C_{1Q}(0)  =  -n_{1,z} n_{2,z}.
\ee 
In the state $\wh{u}_E(\gamma)|[1],1,[2],2 \ra$\/ ,
\be
C_{1Q}(\gamma)  =  \la  [1],1,[2],2 |\,\wh{u}_E\da(\gamma) \,\wh{\xi}\:
\wh{u}_E(\gamma)\,|[1],1,[2],2 \ra .
\ee
Using (\ref{statedef})-(\ref{spin_spec_corr_states}), 
(\ref{uEaction})-(\ref{gopdef}) and (\ref{spincorr}),
\be
C_{1Q}(\gamma)  =  -(1-\sin(2\gamma))n_{1,z} n_{2,z} - \sin(2\gamma)\vec{n}_1 \cdot \vec{n}_2,
\label{spincorrgamma}
\ee
or, for small values of $\gamma$\/,
\be
C_{1Q}(\gamma)_{|\gamma| \ll 1} =  -(1-2\gamma)n_{1,z} n_{2,z} - 2\gamma\vec{n}_1 \cdot\vec{n}_2.
\label{spincorrsmallgamma}
\ee
For $\gamma=\pi/4$\/, 
\be
C_{1Q}(\pi/4) = -\vec{n}_1 \cdot \vec{n}_2,
\ee
the correct value for distinguishable particles in the singlet 
state.\cite{Greenberger_etal90}

For present purposes we are taking $2\gamma$\/, the degree to which the spin
correlation is proportional to  $-\vec{n}_1 \cdot \vec{n}_2$\/,
as a heuristic
measure of entanglement.
This quantity is experimentally measurable, e.g.,
by performing many repetitions of the EPRB experiment with many different
choices of $\vec{n}_1$ \/ and $\vec{n}_2$\/ and determining the
experimental value of $2\gamma$\/ by a least-squares  fit  to the data of the functional
form (\ref{spincorrgamma}) or (\ref{spincorrsmallgamma}).
For a review of entanglement measures in  general, see Ref. 27. % \cite{Horodecki01}.

\subsection{Fock Space Formalism}\label{Fock_section}

\subsubsection{Correspondence with First-Quantized States and Operators}

We consider a Fock space with four annihilation operators, corresponding
to the four possible combinations of spin and species:
\be
\wh{\phi}_{[r]i}, \hspace*{5mm}r,i=1,2, \label{Fspanniops}
\ee
where
\be
\left\{\wh{\phi}_{[r]i}, \wh{\phi}_{[s]j}\right\}=
\left\{\wh{\phi}_{[r]i}\da, \wh{\phi}_{[s]j}\da \right \}=0, \label{Fspphianticomm}
\ee
\be
\left\{\wh{\phi}_{[r]i}, \wh{\phi}_{[s]j}\da \right\}=\delta_{rs} \delta_{ij}.
\ee
Curly brackets denote the anticommutator, $\left\{\wh{A},\wh{B}\right\}=
\wh{A}\wh{B}+\wh{B}\wh{A}$\/. The vacuum (no-particle) state 
 $\sqvk$\/ satisfies
\be
\wh{\phi}_{[r]i}\,\sqvk=0, \hspace*{5mm}r,i=1,2,
\ee
The physical  states of the two-particle Hilbert space correspond to
states created from  $\sqvk$\/ by the creation-operator adjoints of (\ref{Fspanniops}):
\be 
|[r],i,[s],j\ra \leftrightarrow |[r],i,[s],j\rra ,
\ee
where
\be
|[r],i,[s],j\rra = \wh{\phi}_{[r]i}^\dag \,\wh{\phi}_{[s]j}^\dag\, \sqvk . \label{state_from_ops}
\ee
If $\wh{\zeta}_1$\/ is an operator acting in the single-particle Hilbert
space, the Fock space operator $\wh{Z}_1$\/ corresponding to it,
\be
\wh{\zeta}_1 \leftrightarrow \wh{Z}_1,
\ee
is
\be
\wh{Z}_1=\sum_{r,i,s,j}\wh{\phi}_{[r]i}\da\,\zeta_{1,r,i,s,j}\,\wh{\phi}_{[s]j},
\ee
where
\be
\zeta_{1,r,i,s,j}=\la {\cal S}; [r],i|\,\wh{\zeta}_1\, | {\cal S}; [s],j \ra.
\ee
For example, the spin angular momentum is
\be
\wh{\vec{J}}=\frac{1}{2} \sum_{r,i,s,j}\wh{\phi}_{[r]i}\da\,
\la {\cal S}; [r],i|\, \vec{\wh{\sigma}}\, | {\cal S}; [s],j \ra\,
\wh{\phi}_{[s]j}. \label{FspJdef}
\ee
Using (\ref{statedef})-(\ref{complement}), (\ref{Fspphianticomm})-(\ref{state_from_ops}) and (\ref{FspJdef})  we verify that
\begin{eqnarray}
|1,1\rra &  =  & |[1],1,[2],1\rra , \\
|1,0\rra & =  &\frac{1}{\sqrt{2}} \left ( \str  |[1],1,[2],2\rra  + |[1],2,[2],1\rra \right), \\
|0,0\rra & =  & \frac{1}{\sqrt{2}}\left ( \str  |[1],1,[2],2\rra  - |[1],2,[2],1\rra \right), \\
|1,-1\rra &  =  & |[1],2,[2],2\rra ,
\end{eqnarray}
where
\begin{eqnarray}
\wh{\vec{J}} \cdot \wh{\vec{J}}|J,J_z\rra & = & J(J+1)|J,J_z\rra, \\
\wh{J}_z|J,J_z\rra & = & J_z |J,J_z\rra.
\end{eqnarray}

If $\wh{\zeta}_2$\/ is an operator in the two-particle Hilbert-space 
which mediates interactions between two particles,
there is a corresponding
Fock space operator $\wh{Z}_2$\/:
\be
\wh{\zeta_2} \leftrightarrow \wh{Z}_2,
\ee
where
\be
\wh{Z}_2=\frac{1}{2}\sum_{r^\prime,i^\prime,s^\prime,j^\prime}\sum_{r,i,s,j}
\wh{\phi}_{[s\pr]j\pr}^\dag \,\wh{\phi}_{[r\pr]i\pr}^\dag \, 
\zeta_{2,r\pr,i\pr,s\pr,j\pr;r,i,s,j}\,
\wh{\phi}_{[r]i}\, \wh{\phi}_{[s]j},  \label{Z_Fock_space_op}
\ee
and
\be
\zeta_{2,r\pr,i\pr,s\pr,j\pr;r,i,s,j}=
\la{\cal S}^{(1)}{\cal S}^{(2)}; \rho_{[r\pr]}, \alpha_{i\pr}, \rho_{[s\pr]}, \alpha_{j\pr}| 
\, \wh{\zeta}_2 \,
|{\cal S}^{(1)}{\cal S}^{(2)}; \rho_{[r]}, \alpha_{i}, \rho_{[s]}, \alpha_{j}\ra .
\label{zeta_matrix_elt}
\ee

\subsubsection{Time Evolution and Entanglement}

The Fock space operator corresponding to the first quantized operator $\wh{g}$\/
is, using (\ref{Z_Fock_space_op}) and (\ref{zeta_matrix_elt}),
\be
\wh{G}=\frac{1}{2}\sum_{r^\prime,i^\prime,s^\prime,j^\prime}\sum_{r,i,s,j}
\wh{\phi}_{[s\pr]j\pr}^\dag \, \wh{\phi}_{[r\pr]i\pr}^\dag \,
g_{r\pr,i\pr,s\pr,j\pr;r,i,s,j}\,
\wh{\phi}_{[r]i} \, \wh{\phi}_{[s]j}, \label{Gopdef}
\ee
where
\be
g_{r\pr,i\pr,s\pr,j\pr;r,i,s,j} \equiv 
\la{\cal S}^{(1)}{\cal S}^{(2)}; \rho_{[r\pr]} ,\alpha_{i\pr}, \rho_{[s\pr]}, \alpha_{j\pr}| \,
\wh{g}\,
|{\cal S}^{(1)}{\cal S}^{(2)}; \rho_{[r]}, \alpha_{i}, \rho_{[s]}, \alpha_{j}\ra
\ee
Using (\ref{statedef})-(\ref{spin_spec_corr_states}) and (\ref{gopdef}),
\begin{equation} 
\begin{array}{lll}
g_{r\pr,i\pr,s\pr,j\pr;r,i,s,j}&= (i/2)
\left[ \str\left ( \delta_{r\pr 1}\delta_{i\pr 1}\delta_{s\pr 2}\delta_{j\pr 2}
             - \delta_{r\pr 2}\delta_{i\pr 2}\delta_{s\pr 1}\delta_{j\pr 1}\right)
       \left ( \delta_{r 1}\delta_{i 2}\delta_{s 2}\delta_{j 1}
             - \delta_{r 2}\delta_{i 1}\delta_{s 1}\delta_{j 2}\right) \right.&\\
   &  -\left ( \delta_{r\pr, 1}\delta_{i\pr 2}\delta_{s\pr 2}\delta_{j\pr 1}
             - \delta_{r\pr 2}\delta_{i\pr 1}\delta_{s\pr 1}\delta_{j\pr 2}\right)
\left. \left ( \delta_{r 1}\delta_{i 1}\delta_{s 2}\delta_{j 2}
             - \delta_{r 2}\delta_{i 2}\delta_{s 1}\delta_{j 1}\right)\str
\right].&
\end{array} \label{g_components}
\end{equation}
The unitary operator 
\be
\wh{U}_F(\gamma)=\exp(-i\gamma \wh{G})
\ee
acts on two-particle Fock space states in a manner corresponding to the 
action of $\wh{u}_E$\/ on the two-particle physical Hilbert-space
states:
\begin{equation}
\begin{array}{lll}
\wh{U}_F(\gamma)\, \wh{\phi}_{[r]i}\da \, \wh{\phi}_{[s]j}\da\,\sqvk=& &\\
(1/2)\sum_{r\pr i\pr s\pr j\pr}\la[r\pr],i\pr,[s\pr],j\pr |
\, \wh{u}_E(\gamma)\,
|[r],i,[s],j\ra \, 
\wh{\phi}_{[r\pr]i\pr}\da \,\wh{\phi}_{[s\pr]j\pr}\da\,\sqvk.& & \label{U_F_action}
\end{array}
\end{equation}
In particular,
\begin{eqnarray}
\wh{U}_F |[1],2,[2],1\rra & = &  |1,0\rra \\
\wh{U}_F |[1],1,[2],2\rra & = &  |0,0\rra \\
\wh{U}_F |[r],i,[s],j\rra & = &  |[r],i,[s],j\rra, \non \\ 
&&\lla[1],2,[2],1|[r],i,[s],j\rra =\lla[1],1,[2],2|[r],i,[s],j\rra =0.
\end{eqnarray}
where
\be
\wh{U}_F=\wh{U}_F(\pi/4).
\ee

\subsubsection{Spin Correlation}

The Fock space equivalent of the first-quantized operator $\wh{\xi}$\/ is
\be
\wh{\Xi}_F=\frac{1}{2}\sum_{r^\prime,i^\prime,s^\prime,j^\prime}\sum_{r,i,s,j}
\wh{\phi}_{[s\pr]j\pr}^\dag \, \wh{\phi}_{[r\pr]i\pr}^\dag \,
\xi_{r\pr,i\pr,s\pr,j\pr;r,i,s,j}\,
\wh{\phi}_{[r]i} \, \wh{\phi}_{[s]j}, \label{Xi_F_def}
\ee
where
\be
\xi_{r\pr,i\pr,s\pr,j\pr;r,i,s,j} \equiv
\la{\cal S}^{(1)}{\cal S}^{(2)}; \rho_{[r\pr]} ,\alpha_{i\pr} ,\rho_{[s\pr]} ,\alpha_{j\pr}| \,
\wh{\xi} \:
|{\cal S}^{(1)}{\cal S}^{(2)}; \rho_{[r]}, \alpha_{i} ,\rho_{[s]}, \alpha_{j}\ra.
\label{xicomp}
\ee
Using (\ref{statedef}), (\ref{nondegenerate}) and (\ref{spincorr}),
\bea
\xi_{r\pr,i\pr,s\pr,j\pr;r,i,s,j} 
&=&\delta_{r\pr 1} \delta_{s\pr 2}\delta_{r 1} \delta_{s 2}
\la {\cal S}^{(1)};\alpha_{i\pr}|\la {\cal S}^{(2)};\alpha_{j\pr}| (\vec{n}_1\cdot\wh{\vec{\sigma}}^{(1)})(\vec{n}_2\cdot\wh{\vec{\sigma}}^{(2)})
| {\cal S}^{(1)};\alpha_{i} \ra |{\cal S}^{(2)};\alpha_{j}\ra \nonumber \\
&&+\delta_{r\pr 2} \delta_{s\pr 1}\delta_{r 2} \delta_{s 1}
\la {\cal S}^{(1)};\alpha_{i\pr}|\la {\cal S}^{(2)};\alpha_{j\pr}| (\vec{n}_2\cdot\wh{\vec{\sigma}}^{(1)})(\vec{n}_1\cdot\wh{\vec{\sigma}}^{(2)})
| {\cal S}^{(1)};\alpha_{i} \ra |{\cal S}^{(2)};\alpha_{j}\ra. \nonumber \\
 & & \label{xi_val}
\eea
Using (\ref{sigmazaction}), (\ref{alphavalues}), (\ref{sigmaxaction})-(\ref{complement})
and (\ref{xi_val}), (\ref{Xi_F_def}) becomes
\be
\wh{\Xi}_F=\sum_{i\pr, j\pr, i, j}\wh{\phi}_{[2]j\pr}\da \, \wh{\phi}_{[1]i\pr}\da\,
\tilde{\xi}_{i\pr,j\pr,i,j}\,
\wh{\phi}_{[1]i}\, \wh{\phi}_{[2]j} \label{Xi_F}
\ee
where
\bea
\tilde{\xi}_{i\pr,j\pr,i,j}&=&\la {\cal S}^{(1)};\alpha_{i\pr}|\la {\cal S}^{(2)};\alpha_{j\pr}| (\vec{n}_1\cdot\wh{\vec{\sigma}}^{(1)})(\vec{n}_2\cdot\wh{\vec{\sigma}}^{(2)})
| {\cal S}^{(1)};\alpha_{i} \ra |{\cal S}^{(2)};\alpha_{j}\ra \nonumber \\
&=&\left[\str\left(n_{1,x} +in_{1,y}\alpha_i\right)\delta_{i\pr \bar{i}} + n_{1,z} \alpha_i \delta_{i\pr i}\right]
\left[\left(n_{2,x} +in_{2,y}\alpha_j\right)\delta_{j\pr \bar{j}} + n_{2,z} \alpha_j \delta_{j\pr j}\right]. \label{xi_tilde_comp}
\eea

The spin correlation in the state $\wh{U}_F(\gamma)\,|[1],1,[2],2\rra$\/
is
\be
C_F(\gamma) = \lla [1],1,[2],2 |\,\wh{U}_F\da(\gamma)\, \wh{\Xi}_F \, \wh{U}_F(\gamma)\,|[1],1,[2],2\rra
\label{C_F_def} 
\ee
Using (\ref{spin_species}), (\ref{uEdef})-(\ref{gopdef}), (\ref{state_from_ops}), (\ref{g_components}),
(\ref{U_F_action}), %(\ref{xi_tilde_comp}) 
and (\ref{Xi_F})-(\ref{C_F_def}), we find that,
in agreement with the first-quantized results (\ref{spincorrgamma}), (\ref{spincorrsmallgamma}),
\be
C_F(\gamma)= -(1-\sin(2\gamma))n_{1,z} n_{2,z} - \sin(2\gamma)\vec{n}_1 \cdot \vec{n}_2
\ee
or, for small $\gamma$\/,
\be
C_F(\gamma)_{|\gamma| \ll 1} =  -(1-2\gamma)n_{1,z} n_{2,z} - 2\gamma\vec{n}_1 \cdot\vec{n}_2.
\ee

\section{EPRB ENTANGLEMENT IN QUANTUM FIELD THEORY}

\subsection{Operators and States}

We now consider a quantum field theory, in three space dimensions, of
a nonrelativistic spin 1/2 particle which comes in two distinct species.
We work from the outset in the Heisenberg picture. The dynamical variables of the
theory are  time-dependent field operators defined at each point in
space, $\wh{\phi}_{[r]i}(\vec{x},t)$\/, $r,i=1,2$\/.
At the initial time $t_0$\/ these operators satisfy 
\be
\left\{\wh{\phi}_{[r]i}(\vec{x}), \wh{\phi}_{[s]j}(\vec{y})\right\}=
\left\{\wh{\phi}_{[r]i}\da(\vec{x}), \wh{\phi}_{[s]j}\da(\vec{y}) \right \}=0,
\label{anticom1}
\ee
\be
\left\{\wh{\phi}_{[r]i}(\vec{x}), \wh{\phi}_{[s]j}\da(\vec{y}) \right\}=\delta_{rs} \delta_{ij}\delta^3(\vec{x}-\vec{y}),
\label{anticom2}
\ee
where
\be
\wh{\phi}_{[r]i}(\vec{x})\equiv \wh{\phi}_{[r]i}(\vec{x},t_0),
\ee
\be
\wh{\phi}_{[r]i}\da(\vec{x})\equiv \wh{\phi}_{[r]i}\da(\vec{x},t_0).
\ee
By virtue of unitary time evolution, the equal-time anticommutation relations
(\ref{anticom1}) and (\ref{anticom2}) are also satisfied at all later
times $t$\/:
\be
\left\{\wh{\phi}_{[r]i}(\vec{x},t), \wh{\phi}_{[s]j}(\vec{y},t)\right\}=
\left\{\wh{\phi}_{[r]i}\da(\vec{x},t), \wh{\phi}_{[s]j}\da(\vec{y},t) \right \}=0,
\label{anticom1t}
\ee
\be
\left\{\wh{\phi}_{[r]i}(\vec{x},t), \wh{\phi}_{[s]j}\da(\vec{y},t) \right\}=\delta_{rs} \delta_{ij}\delta^3(\vec{x}-\vec{y}),
\label{anticom2t}
\ee

The time-dependent Heisenberg-picture field operators act on
time-independent states defined at $t=t_0$\/.  For the vacuum 
state at $t_0$\/ we will use the same symbol as for the vacuum state of
the  Fock space of  Sec. \ref{Fock_section}, $\sqvk$\/:
\be
\wh{\phi}_{[r]i}(\vec{x})\,\sqvk=0, \hspace*{5mm}r,i=1,2, \label{vacanni}
\ee
An arbitrary normalized two-particle Heisenberg-picture state can be written as
\be
|\psi\rra = \sum_{r,i,s,j} \int d^3\vec{x}\, d^3\vec{y}\,
\psi_{[r]i[s]j}(\vec{x},\vec{y})\,\wh{\phi}_{[r]i}\da(\vec{x})\,\wh{\phi}_{[s]j}\da(\vec{y})\,
\sqvk
\ee
where $\psi_{[r]i[s]j}(\vec{x},\vec{y})$\/ is a complex c-number function satisfying
the normalization condition
\be
\sum_{r,i,s,j} \int d^3\vec{x} \, d^3\vec{y}\,
\psi_{[r]i[s]j}^\ast(\vec{x},\vec{y})\, \psi_{[r]i[s]j}(\vec{x},\vec{y})
=1/2
\ee
and, without loss of generality, the antisymmetry condition
\be
\psi_{[r]i[s]j}(\vec{x},\vec{y})=-\psi_{[s]j[r]i}(\vec{y},\vec{x}).
\ee

\subsection{Hamiltonian and Exact Equation of Motion}

The Hamiltonian is the sum  of two parts,
\be
\wh{H}=\wh{H}_0+ \eps\wh{H}_1  \label{Hamsum}
\ee
where $\eps$\/ is a small parameter.
The free Hamiltonian is
\be
\wh{H}_0=\sum_{r,i}\int d^3 \vec{x} \,
\wh{\phi}_{[r]i}\da(\vec{x},t)
\left(\frac{-\vec{\nabla}^2}{2m}\right)
\wh{\phi}_{[r]i}(\vec{x},t).                 \label{Hfree}
\ee
For the interaction Hamiltonian we choose an interaction which, at each
point $\vec{x},$\/ has the same form as the entanglement generator (\ref{Gopdef}) of
the Fock space of Sec. 
\ref{Fock_section}:
\be
\wh{H}_1=\frac{1}{2}\sum_{r\pr,i\pr,s\pr,j\pr}\sum_{r,i,s,j}\int d^3 \vec{x}\, \kappa(\vec{x},t)
\,\wh{\phi}_{[s\pr]j\pr}\da(\vec{x},t)\,\wh{\phi}_{[r\pr]i\pr}\da(\vec{x},t)\,
g_{r\pr,i\pr,s\pr,j\pr;r,i,s,j}\,
\wh{\phi}_{[r]i}(\vec{x},t)\,\wh{\phi}_{[s]j}(\vec{x},t).   \label{Hint}
\ee
The possibility of spacetime dependence in the coupling, $\ka(\vec{x},t)$\/, has been allowed for.

Field operators evolve in time according to the Heisenberg equation of
motion
\be
\frac{\partial}{\partial t} \wh{\phi}_{[r]i}(\vec{x},t)
=i[\wh{H},\wh\phi_{[r]i}(\vec{x},t)],                          \label{Heqmo}
\ee
where we take $\hbar=1$\/.
Using (\ref{anticom1t}), (\ref{anticom2t}), 
and (\ref{Hamsum})-(\ref{Hint})
in (\ref{Heqmo}), we obtain the exact equation of motion for $\wh\phi_{[r]i}(\vec{x},t)$\/,
\be
\frac{\partial}{\partial t} \wh{\phi}_{[r]i}(\vec{x},t)
=i\frac{\vec{\nabla}^2}{2m}\wh{\phi}_{[r]i}(\vec{x},t)
+i\eps \kappa(\vec{x},t)\sum_{s\pr, j\pr}\sum_{q,k,s,j}
g_{r,i,s\pr,j\pr;q,k,s,j}\,
\wh{\phi}_{[s\pr]j\pr}\da(\vec{x},t)\,\wh{\phi}_{[q]k}(\vec{x},t)\,\wh{\phi}_{[s]j}(\vec{x},t).
\label{exacteqmo}
\ee

\subsection{First-order Perturbation Theory}

We obtain a solution to eq. (\ref{exacteqmo}) 
for $\wh{\phi}_{[r]i}(\vec{x},t)$\/ using a straightforward
perturbative approach (see, e.g., Ref. 28) %\cite{Kallen})
in terms of the small parameter $\eps$\/.
We look for a solution  of the form
\be
\wh{\phi}_{[r]i}(\vec{x},t)=\wh{\phi}_{0,[r]i}(\vec{x},t)+\eps \wh{\phi}_{1,[r]i}(\vec{x},t).
\label{pertexp}
\ee
Using (\ref{pertexp}) in (\ref{exacteqmo}) we  obtain the zeroth-order equation of motion
\be
i\frac{\partial}{\partial t} \wh{\phi}_{0,[r]i}(\vec{x},t)=
-\frac{\vec{\nabla}^2}{2m}\wh{\phi}_{0,[r]i}(\vec{x},t)
\ee
and the first-order equation of motion
\be
i\frac{\partial}{\partial t} \wh{\phi}_{1,[r]i}(\vec{x},t)=
-\frac{\vec{\nabla}^2}{2m}\wh{\phi}_{1,[r]i}(\vec{x},t)
+\wh{J}_{[r]i}(\vec{x},t), 
\ee
where
\be
\wh{J}_{[r]i}(\vec{x},t)=-\kappa(\vec{x},t)
\sum_{s\pr,j\pr}\sum_{q,k,s,j}g_{r,i,s\pr,j\pr;q,k,s,j}\,
\wh{\phi}_{0,[s\pr]j\pr}\da(\vec{x},t)\,\wh{\phi}_{0,[q]k}(\vec{x},t)\,
\wh{\phi}_{0,[s]j}(\vec{x},t).
\ee
We impose the initial condition
\be
\wh{\phi}_{1,[r]i}(\vec{x},t_0)=0,
\ee
so
\be
\wh{\phi}_{0,[r]i}(\vec{x},t_0)=\wh{\phi}_{[r]i}(\vec{x},t_0)=\wh{\phi}_{[r]i}(\vec{x}).
\ee
The solution for $\wh{\phi}_{[r]i}(\vec{x},t)$\/ in terms of 
field operators at the initial time $t_0$\/ is
\be
\wh{\phi}_{[r]i}(\vec{x},t)=
\int d^3 \vec{x}\pr G(\vec{x}-\vec{x}\pr,t-t_0)\wh{\phi}_{[r]i}(\vec{x})
-i\eps\int_{t_0}^t dt\pr \int d^3\vec{x}\pr G(\vec{x}-\vec{x}\pr,t-t')
\wh{J}_{[r]i}(\vec{x}\pr,t\pr), \label{pertsol}
\ee
where
\be
\begin{array}{rl}
\wh{J}_{[r]i}(\vec{x},t)& = 
-\ka(\ve{x},t)\sum_{s\pr,j\pr}\sum_{q,k,s,j}g_{r,i,s\pr,j\pr;q,k,s,j}
\int d^3\ve{z}\, d^3\ve{z}\pr\, d^3\ve{z}\ppr \\
& G^\ast(\ve{x}-\ve{z},t-t_0) G(\ve{x}-\ve{z}\pr,t-t_0)G(\ve{x}-\ve{z}\ppr,t-t_0)  \,
\wh{\phi}_{[s\pr]j\pr}\da(\vec{z})\,\wh{\phi}_{[q]k}(\vec{z\pr})\,
\wh{\phi}_{[s]j}(\vec{z\ppr}).
\end{array} \label{Jpertsol}
\ee
and $G(\ve{x}-\ve{x}\pr,t-t\pr)$\/ is the Schr\"{o}dinger Green's function
\be
G(\ve{x}-\ve{x}\pr,t-t\pr)=
\left(\frac{-2mi}{4\pi(t-t\pr)}\right)^{3/2}
\exp \left( \frac{im|\ve{x}-\ve{x}\pr|^2}{2(t-t\pr)} \right), \label{Gxt}
\ee
satisfying
\be
\left(i\frac{\partial}{\partial t}+\frac{\vec{\nabla}^2}{2m}\right)G(\ve{x}-\ve{x}\pr,t-t\pr)
=0
\ee
and
\be
G(\ve{x}-\ve{x}\pr,0)=\delta^3(\ve{x}-\ve{x}\pr). \label{Gx0}
\ee
From (\ref{Gxt}) it follows that 
\bea
G(\ve{x}\pr-\ve{x},t-t\pr)&=&G(\ve{x}-\ve{x}\pr,t-t\pr), \label{Gsymx}\\
G(\ve{x}-\ve{x}\pr,t\pr-t)&=&G\as(\ve{x}-\ve{x}\pr,t-t\pr), \label{Gsymt}
\eea
and
\be
\int d^3\ve{x}\ppr G(\ve{x}-\ve{x}\ppr,t-t\ppr)G\as(\ve{x}\pr-\ve{x}\ppr,t\pr-t\ppr)
=G(\ve{x}-\ve{x}\pr,t-t\pr). \label{Gconv}
\ee

\subsection{Spin Correlation}

For the initial-time Heisenberg-picture state we take a state,
which we denote $|\psi_0\rra,$\/ containing
one spin-up particle of species 1 and one spin-down particle of
species 2:
\bea
\psi_{0,[1]1[2]2}(\vec{x},\vec{y})&=&-\psi_{0,[2]2[1]1}(\vec{y},\vec{x})
=\frac{1}{2}\,\psi_{[1]1}(\vec{x})\,\psi_{[2]2}(\vec{y}), \non \\
\psi_{0,[r]i[s]j}(\vec{x},\vec{y})&=&0,  \hspace*{5mm} r,i,s,j \neq 1,1,2,2 
\; \mbox{\rm or } 2,2,1,1,
\eea
where
\be
\int d^3 \ve{x} \,\psi_{[1]1}\as(\ve{x})\,\psi_{[1]1}(\ve{x})=
\int d^3 \ve{x}\, \psi_{[2]2}\as(\ve{x})\,\psi_{[2]2}(\ve{x})=1.
\ee
That is,
\be
|\psi_0\rra=\int d^3 \ve{x}\, d^3 \ve{y}\,\psi_{[1]1}(\ve{x})\,\psi_{[2]2}(\ve{y})\,
\wh{\phi}_{[1]1}\da(\vec{x})\, \wh{\phi}_{[2]2}\da(\vec{y})\, \sqvk. \label{psi0}
\ee

The field-theoretic operator corresponding to the first-quantized
spin-correlation operator (\ref{spincorr}) is
\bea
\wh{\Xi}(t)&=&\frac{1}{2}\int d^3 \ve{x\pr}\, d^3 \ve{y\pr}\, d^3 \ve{x} \,d^3 \ve{y}\,
\sum_{r^\prime,i^\prime,s^\prime,j^\prime}\sum_{r,i,s,j} \non \\
&&\wh{\phi}_{[s\pr]j\pr}^\dag(\vec{y\pr},t)\, \wh{\phi}_{[r\pr]i\pr}^\dag (\vec{x\pr},t)\,
\xi_{r\pr,i\pr,s\pr,j\pr;r,i,s,j}(\ve{x\pr},\ve{y\pr},\ve{x},\ve{y})\,
\wh{\phi}_{[r]i}(\vec{x},t)\, \wh{\phi}_{[s]j}(\vec{y},t), \label{Xi_def}
\eea
where
\be
\xi_{r\pr,i\pr,s\pr,j\pr;r,i,s,j}(\ve{x}\pr,\ve{y}\pr,\ve{x},\ve{y})=
\la{\cal S}^{(1)}{\cal S}^{(2)}; \rho_{[r\pr]}, \alpha_{i\pr},\ve{x}\pr,
 \rho_{[s\pr]}, \alpha_{j\pr},\ve{y}\pr| 
\, \wh{\xi} \,
|{\cal S}^{(1)}{\cal S}^{(2)}; \rho_{[r]}, \alpha_{i},\ve{x},
 \rho_{[s]}, \alpha_{j}, \ve{y} \ra.
\ee
Since $\wh{\xi}$\/ acts trivially on spatial coordinates,
\be
\xi_{r\pr,i\pr,s\pr,j\pr;r,i,s,j}(\ve{x}\pr,\ve{y}\pr,\ve{x},\ve{y})=
\xi_{r\pr,i\pr,s\pr,j\pr;r,i,s,j}\delta^3(\ve{x}\pr - \ve{x})\delta^3(\ve{y}\pr - \ve{y}),
\ee
where $\xi_{r\pr,i\pr,s\pr,j\pr;r,i,s,j}$\/ is as given in (\ref{xi_val}).
Using  
(\ref{xi_val}) and (\ref{xi_tilde_comp}),  (\ref{Xi_def}) becomes
\be
\wh{\Xi}(t)=\int d^3 \ve{x}\, d^3 \ve{y}\,\sum_{i\pr, j\pr, i, j}
\wh{\phi}_{[2]j\pr}\da(\ve{y},t)\, \wh{\phi}_{[1]i\pr}\da(\ve{x},t)\,
\tilde{\xi}_{i\pr,j\pr,i,j}\,
\wh{\phi}_{[1]i}(\ve{x},t)\, \wh{\phi}_{[2]j}(\ve{y},t). \label{Xi_val}
\ee
The spin correlation at time $t$\/ given the initial state $|\psi_0\rra$\/ is
\be 
C(t)=\lla \psi_0 | \, \wh{\Xi}(t)\, | \psi_0 \rra. \label{Cqftdef}
\ee
Using (\ref{g_components}), (\ref{xi_tilde_comp}), (\ref{anticom1}), (\ref{anticom2}), (\ref{vacanni}), (\ref{pertsol}), (\ref{Jpertsol}),  (\ref{Gx0})-(\ref{Gconv}),
(\ref{psi0}) and (\ref{Xi_val}) we find
\be
C(t)=-(1-\eps L(t))n_{1,z}n_{2,z}-\eps L(t)\ve{n}_1 \cdot \ve{n}_2  \label{Cqft}
\ee
where
\be
\begin{array}{ll}
L(t)=&2\int_{t_0}^t dt\ppr \int d^3 \ve{x}\ppr \ka(\ve{x}\ppr,t\ppr)\\
&\left[ \int d^3 \ve{y}\pr \,\psi_{[2]2}\as(\ve{y}\pr)\, G\as(\ve{y}\pr-\ve{x}\ppr,t\ppr-t_0)\right]
\left[ \int d^3 \ve{x}\pr \,\psi_{[1]1}\as(\ve{x}\pr)\, G\as(\ve{x}\pr-\ve{x}\ppr,t\ppr-t_0)\right]\\
&\left[ \int d^3 \ve{x}\, \psi_{[1]1}(\ve{x})\, G(\ve{x}-\ve{x}\ppr,t\ppr-t_0)\right]
\left[ \int d^3 \ve{y}\, \psi_{[2]2}(\ve{y}) \,G(\ve{y}-\ve{x}\ppr,t\ppr-t_0)\right].
\end{array} \label{Ldef}
\ee

\subsection{Gaussian Wavepackets}

Now suppose that the functions $\psi_{[1]1}(\ve{x}),$\/ $\psi_{[2]2}(\ve{x}),$\/
are of the form
\bea
\psi_{[1]1}(\ve{x})&=&\psi_g(\ve{x}_1,\ve{v}_1;\ve{x}) \label{psig11} \\
\psi_{[2]2}(\ve{x})&=&\psi_g(\ve{x}_2,\ve{v}_2;\ve{x}) \label{psig22}
\eea
where
\be
\psi_g(\ve{x}_i,\ve{v}_i;\ve{x})=\lf( \frac{\alpha}{\pi}\rt)^{3/4}
\exp\lf(-\frac{\al|\ve{x}-\ve{x}_i|^2}{2} +im\ve{v}_i\cdot (\ve{x}-\ve{x}_i)\rt).
\label{psigdef}
\ee
That is, the initial state consists of two Gaussian wavepackets each of which,
at time $t_0$\/, has width
$1/\sqrt{\al}$\/, the wavepacket for the particle of species $i$\/ being  centered at position $\ve{x}_i$\/ and moving with velocity $\ve{v}_i$\/.
Using (\ref{Gxt}) and (\ref{psig11})-(\ref{psigdef}), $L(t)$\/ in eq. (\ref{Ldef})  in this
case has the value
\be
L_g(t)=2\int_{t_0}^t dt\ppr \lf(\frac{A(t\ppr)}{\pi}\rt)^3
\int d^3 \ve{x}\ppr \ka(\ve{x}\ppr,t\ppr)
\exp\lf(-A(t\ppr) \lf[|\ve{x}\ppr-\ve{c}_1(t\ppr)|^2 + |\ve{x}\ppr-\ve{c}_2(t\ppr)|^2\rt]\rt),
\label{Lg}
\ee
where
\be
A(t\ppr)=\frac{\al}{1+\al^2\lf(t\ppr-t_0\rt)^2/m^2} \label{Adef}
\ee
and
\be
\vec{c}_i(t\ppr)=\vec{x}_i + \ve{v}_i(t\ppr-t_0), \hspace*{5mm} i=1,2.  \label{cdef}
\ee
$\vec{c}_i(t\ppr)$\/ is the trajectory of the center of the wavepacket of
the particle of species $i$\/.

\subsubsection{Point Interaction}

Suppose that the entangling interaction between particles of different
species is only ``turned on'' in a very limited region of space, idealized as
a point $\ve{x}_I$\/, and only for a
brief extent of time idealized as a single moment $t_I$\/:
\be
\ka(\vec{x},t\ppr)=\ka \delta^3(\ve{x}-\ve{x}_I)\delta(t\ppr-t_I).
\ee 
$L_g(t)$\/ in (\ref{Lg}) then has the value
\be
L_{g-PI}(t)=2  \ka \lf(\frac{A(t_I)}{\pi}\rt)^3
\exp\lf(-A(t_I) \lf[|\ve{x}_I-\ve{c}_1(t_I)|^2 + 
|\ve{x}_I-\ve{c}_2(t_I)|^2\rt]\rt).
\ee
This quantity will be exponentially small unless 
\be
\ve{c}_1(t_I) \approx \ve{c}_2(t_I) \approx \ve{x}_I.
\ee
That is,  there will be little entanglement generated
unless the initial locations and velocities of the
centers of the wavepackets are such as to arrive
near the point $\ve{x}_I$\/ very close to the time
$t_I$\/.

\subsubsection{Position-Independent Interaction}

Consider now that the coupling $\ka(\ve{x},t\ppr)$\/ is
independent of position but still possibly time-dependent:
\be
\ka(\vec{x},t\ppr)=\ka(t\ppr).
\ee
Using this in (\ref{Lg}), $L_g(t)$\/ becomes
\be
L_{g-PII}=\frac{1}{\sqrt 2}\int_{t_0}^t dt\ppr \ka(t\ppr)
\lf(\frac{A(t\ppr)}{\pi}\rt)^{3/2}
\exp\lf(-\frac{1}{2}A(t\ppr)|\ve{c}_1(t\ppr)-\ve{c}_2(t\ppr)|^2 \rt)
\label{Lgpii}
\ee
We can approximate the integral in (\ref{Lgpii})
by the method of steepest descent (see, e.g., Ref. 29), %\cite{KornKorn}),
\be
\int dt\ppr \, f(t\ppr) \approx \sqrt\frac{2\pi f(t_c)^3}{-\ddot{f}(t_c)},
\ee
where $t_c$\/ is the point at which the integrand $f(t\ppr)$\/
attains its maximum ($\dot{f}(t_c)=0$\/).
The approximation is better the more the integrand peaks at
one value of the integration variable. Therefore, taking into
account (\ref{Adef}) and (\ref{Lgpii}), we consider the limit
in which the width $1/\sqrt{\al}$\/ of each wavepacket becomes small and the mass $m$\/ of each particle becomes large in such a way that
\be
\frac{\al^2(t-t_0)^2}{m^2} \rightarrow 0.
\ee
In this limit 
\be
A(t\ppr) \approx \al
\ee
so
\be
L(t)_{g-PII} \approx \frac{1}{\sqrt 2}\lf(\frac{\al}{\pi}\rt)^{3/2}
\int_{t_0}^t dt\ppr {\cal L}(t\ppr)
\ee
where
\be
{\cal L}(t\ppr)=\ka(t\ppr)\exp\lf(-\frac{\al}{2} |\ve{c}_1(t\ppr)-\ve{c}_2(t\ppr)|^2\rt). \label{calLdef}
\ee
Setting $\dot{\cal L}(t_c)=0$\/ we find, using (\ref{cdef}) and (\ref{calLdef}),
\be
t_c=t_0 + \frac{1}{\al |\ve{v}_1-\ve{v}_2|^2}\lf[\frac{\dot{\ka}(t_c)}{\ka(t_c)}
-\al (\ve{v}_1-\ve{v}_2)\cdot (\ve{x}_1-\ve{x}_2)\rt]
\ee
If the coupling $\ka(t\ppr)$\/ changes sufficiently smoothly with time in the
vicinity of $t_c$\/---specifically,
\be
|\dot{\ka}(t_c)/\ka(t_c)| \ll \al(\ve{v}_1-\ve{v}_2)\cdot (\ve{x}_1-\ve{x}_2)
\ee
and
\be
|\ddot{\ka}(t_c)/\ka(t_c)| \ll \al|\ve{v}_1-\ve{v}_2|^2
\ee
---then 
\be
t_c \approx t_{min} =t_0-\frac{(\ve{v}_1-\ve{v}_2)\cdot (\ve{x}_1-\ve{x}_2)}
{|\ve{v}_1-\ve{v}_2|^2}
\ee
where $t_{min}$\/ is the time at which the centers of the two wavepackets attain
their minimum separation $d_{min}$\/,
\be
d_{min}=\min_{t\ppr} |\ve{c}_1(t\ppr)-\ve{c}_2(t\ppr)|,
\ee
\be
t_{min}=\arg\min |\ve{c}_1(t\ppr)-\ve{c}_2(t\ppr)|,
\ee
and the degree of entanglement at time $t$\/ is 
\be
L(t)_{g-PII} \approx \frac{\al\ka(t_{min})}{\pi}\sqrt\frac{1}{|\ve{v}_1-\ve{v}_2|^2}
\exp\lf(-\frac{\al}{2}d_{min}^2\rt).
\ee
This result is again intuitively reasonable.  The degree of entanglement
at the final time $t$ is an exponentially-decreasing function of the ``miss distance''
of the centers of the two Gaussian wavepackets,  
and  is inversely proportional to the 
speed $|\ve{v}_1-\ve{v}_2|$ with which they pass each other
(lower speed providing more time to interact while the wavepackets 
most nearly overlap.)

\section{THE VACUUM REPRESENTATION}\label{vacrepsec}

At first glance, the results of the previous section for the expected values of the spin
correlation $C(t)$\/ may well seem to fit with a conceptual picture of entanglement being
generated and then carried along to measuring devices in a local manner, as operator-valued
wavepackets encounter each other and then proceed on to the experimenters'
apparatus.  But, in fact, the formalism presented above nowhere displays 
a local mechanism for transporting the information $\ve{x}_i,$\/ $\ve{v}_i,$\/
about the initial conditions at time $t_0$\/ to later times. Whether we examine the exact
equation of motion (\ref{exacteqmo}) for $\wh{\phi}(\ve{x},t)$ or the perturbative
solution (\ref{pertsol}), (\ref{Jpertsol}), we see that---despite the fact that
 $\wh{\phi}(\ve{x},t)$\/ is  expressed in the latter in terms of the 
time-$t_0$\/ operators
$\wh{\phi}(\ve{x})$\/, $\wh{\phi}\da(\ve{x})$\/---$\wh{\phi}(\ve{x},t)$\/  is completely independent,
at any time $t$\/, of the information contained in the initial state $|\psi_0\rra$\/.
 
This result is due to, and serves to illustrate,
the respective roles assigned by the Heisenberg picture formalism
to operators and to states. 
In the Schr\"{o}dinger picture, information on initial conditions and
time evolution is carried by the state vector. In the Heisenberg picture,
information on time evolution is transfered to the operators, but information
on initial conditions---{\em physical}\/ initial conditions---continues to reside 
in the  state vector.
E.g., in the case at hand, the  condition (\ref{vacanni}) imposed  on
the $t=t_0$\/ field operators $\wh{\phi}_{[r]i}(\vec{x})$\/
imparts to them no information on where one is more or less likely to find a
particle. Therefore, the operators
$\wh{\phi}_{[r]i}(\vec{x},t)$\/ at later times are also lacking
this information; it can only be obtained by also using the 
state vector $|\psi_0\rra$\/.  
``As a Heisenberg operator evolves, at each time it still 
corresponds to its full spectrum of values, the same spectrum
at all times.  In other words, a Heisenberg operator does not correspond to what is,
at one time, a single value of a physical quantity which value then
changes over time...Only the Heisenberg
operator {\em plus}\/ a time-independent state picks out a specific value.''\cite{Teller95}
In the usual Heisenberg-picture quantum field theory, this results in the situation
that ``the quantized solution to a field equation is not {\em a}\/ solution, it
is in some sense {\em the}\/ solution \ldots A `quantum field' corresponds more
closely to  a `general solution' to a field equation than to a specific solution
reflecting complete initial and boundary values. As the analogue of a classical
`general solution,' we again describe the `operator-valued quantum field' most accurately
as a field determinable, something that charts the spatiotemporal relations of
any of a large set of possible field configurations.''\cite{Teller95}
(See also Ref. 31, 
%{\cite{DeWitt72}, 
p. 182.)

What is necessary to see explicitly the locality of Heisenberg-picture quantum field
theory is to complete the transfer of information from states to operators.
As pointed out by Deutsch and Hayden\cite{DeutschHayden00} in the context of Heisenberg-picture
quantum mechanics, this can be accomplished by performing a unitary transformation
which takes the Heisenberg-picture state vector, containing the initial-condition
information, to some standard state vector which is the same regardless of
initial conditions.  To keep the physics unchanged, the same unitary transformation
must of course be applied to the operators.  In such a representation, 
``the term `state vector' becomes a misnomer, for [it] contains no information \ldots
All the information is contained in the observables.''\cite{DeutschHayden00}

For field theory a natural choice for the standard state is the vacuum state $\sqvk$\/.
Consider the operator 
\be
\wh{V}[\psi_{[1]1},\psi_{[2]2}](\theta) \equiv \exp(\theta \wh{W}[\psi_{[1]1},\psi_{[2]2}]),
\label{Vthetadef}
\ee
$\theta$\/ real, where 
\be
\begin{array}{l}
\wh{W}[\psi_{[1]1},\psi_{[2]2}] \equiv \\
\hspace*{.5cm} \int d^3 \ve{x} \, d^3 \ve{y}\, \lf(\psi_{[1]1}(\ve{x})\,\psi_{[2]2}(\ve{y})\,
\wh{\phi}_{[1]1}\da(\vec{x})\, \wh{\phi}_{[2]2}\da(\vec{y})-
\psi_{[1]1}\as(\ve{x})\, \psi_{[2]2}\as(\ve{y})\,
\wh{\phi}_{[2]2}(\vec{y})\,\wh{\phi}_{[1]1}(\vec{x})\rt).   
\end{array}
\label{Wdef}
\ee
The square brackets here denote functional dependence; for notational convenience
this will not be explicitly indicated below. $\wh{W}$\/ is skew-Hermitian, so
$\wh{V}(\theta)$ is unitary. 
Using (\ref{anticom1}), (\ref{anticom2}), (\ref{vacanni}), (\ref{Wdef}) and mathematical induction, we
find that
\be
\wh{W}^{2n}\sqvk = (-1)^n\sqvk, \hspace*{5mm} n=0,1,2, \ldots,
\ee
\be
\wh{W}^{2n+1}\sqvk = (-1)^n |\psi_0\rra \hspace*{5mm} n=0,1,2, \ldots,
\ee
so
\be
\wh{V}(\theta)\sqvk=\cos(\theta)\sqvk + \sin(\theta) | \psi_0\rra.
\ee
Defining
\be
\wh{V}=\wh{V}(\pi/2) \label{Vdef},
\ee
we see that 
\be
\wh{V}\sqvk = |\psi_0\rra , 
\ee
\be
\wh{V}\da|\psi_0\rra = \sqvk.
\ee

We can therefore transform to a ``vacuum representation'' in which 
the Heisenberg-picture state vector is simply $\sqvk$\/ and the 
initial-time operators, $\wh{\phi}_{V, [r]i}(\ve{x})$\/, depend, through
$\wh{V}$\/, on the information which, in the
usual representation, is contained in $|\psi_0\rra$\/. That is,
\be
\wh{\phi}_{V, [r]i}(\ve{x})=\wh{V}\da\,\wh{\phi}_{[r]i}(\ve{x})\,\wh{V}. \label{phiVdef}
\ee
Using (\ref{Vthetadef}), (\ref{Vdef}) and (\ref{phiVdef})
\begin{displaymath}
\wh{\phi}_{V, [r]i}(\ve{x})= 
\wh{\phi}_{[r]i}(\ve{x}) + \frac{\pi}{2}[\wh{\phi}_{[r]i}(\ve{x}),\wh{W}]
+\frac{1}{2!}\lf(\frac{\pi}{2}\rt)^2[[\wh{\phi}_{[r]i}(\ve{x}),\wh{W}],\wh{W}] 
\end{displaymath}
\be
\hspace*{1in} +\frac{1}{3!}\lf(\frac{\pi}{2}\rt)^3
[[[\wh{\phi}_{[r]i}(\ve{x}),\wh{W}],\wh{W}],\wh{W}] + \ldots \ .
\label{phiVexp}
\ee
(See, e.g., Ref. 32.) %\cite{Veltman94}.)
Brackets denote the commutator, $[\wh{A},\wh{B}]=
\wh{A}\wh{B}-\wh{B}\wh{A}$\/.
 Using (\ref{anticom1}), (\ref{anticom2})
and (\ref{Wdef}),
\be
[\wh{\phi}_{[r]i}(\ve{x}),\wh{W}]=
\delta_{r1}\delta_{i1}\psi_{[1]1}(\ve{x})\int d^3\ve{y} \psi_{[2]2}(\ve{y})\wh{\phi}_{[2]2}\da(\ve{y})
-\delta_{r2}\delta_{i2}\psi_{[2]2}(\ve{x})\int d^3\ve{y} \psi_{[1]1}(\ve{y})\wh{\phi}_{[1]1}\da(\ve{y}). \label{phiWcom}
\ee
From (\ref{phiVexp}) and (\ref{phiWcom}), we see that $\wh{\phi}_{V, [r]i}(\ve{x})$\/
is identical to $\wh{\phi}_{[r]i}(\ve{x})$\/ at any location $\ve{x}$\/
where 
$\psi_{[r]i}(\ve{x})$\/   vanishes:
\be
\wh{\phi}_{V, [r]i}(\ve{x})=\wh{\phi}_{[r]i}(\ve{x}),\hspace*{5mm} \ve{x} \notin 
\mbox{\rm support of} \ \psi_{[r]i}.
\ee

Since $\wh{V}$\/ is independent of time, the equation of motion (\ref{exacteqmo})
retains its form 
\begin{displaymath}
\hspace*{-2in}\frac{\partial}{\partial t} \wh{\phi}_{V,[r]i}(\vec{x},t)
= i\frac{\vec{\nabla}^2}{2m}\wh{\phi}_{V,[r]i}(\vec{x},t) 
\end{displaymath}
\be
 \hspace*{1in} +i\eps \kappa(\vec{x},t)\sum_{s\pr, j\pr}\sum_{q,k,s,j}
g_{r,i,s\pr,j\pr;q,k,s,j}\,
\wh{\phi}_{V,[s\pr]j\pr}\da(\vec{x},t)\,\wh{\phi}_{V,[q]k}(\vec{x},t)\,
\wh{\phi}_{V,[s]j}(\vec{x},t),
\label{Vexacteqmo}
\ee
as does the first-order perturbative solution (\ref{pertsol}), (\ref{Jpertsol}):
\bea
\wh{\phi}_{V,[r]i}(\vec{x},t)& = &
\int d^3 \vec{x}\pr G(\vec{x}-\vec{x}\pr,t-t_0)\wh{\phi}_{V,[r]i}(\vec{x}) \nonumber \\
&&-i\eps\int_{t_0}^t dt\pr \int d^3\vec{x}\pr G(\vec{x}-\vec{x}\pr,t-t')
\wh{J}_{V,[r]i}(\vec{x}\pr,t\pr), \label{Vpertsol}
\eea
\be
\begin{array}{l}
\wh{J}_{V,[r]i}(\vec{x},t) =
-\ka(\ve{x},t)\sum_{s\pr,j\pr}\sum_{q,k,s,j}g_{r,i,s\pr,j\pr;q,k,s,j}
\int d^3\ve{z}\, d^3\ve{z}\pr \, d^3\ve{z}\ppr \nonumber \\
\; \; \; \; \; \; G^\ast(\ve{x}-\ve{z},t-t_0)G(\ve{x}-\ve{z}\pr,t-t_0)G(\ve{x}-\ve{z}\ppr,t-t_0)  
\, \wh{\phi}_{V,[s\pr]j\pr}\da(\vec{z})\, \wh{\phi}_{V,[q]k}(\vec{z\pr})\,
\wh{\phi}_{V,[s]j}(\vec{z\ppr}),
\end{array}
\label{VJpertsol}
\ee
where, of course,
\be
\wh{\phi}_{V, [r]i}\da(\ve{x},t)=\wh{V}\da\,\wh{\phi}_{[r]i}\da(\ve{x},t)\,\wh{V},
\ee
\be
\wh{J}_{V, [r]i}(\ve{x},t)=\wh{V}\da\,\wh{J}_{[r]i}(\ve{x},t)\,\wh{V},
\ee
etc.  

In these expressions the manner in which information is encoded in
the operators at $t_0$\/ and subsequently transferred, in accordance with
the rules of a local differential equation, to operators at later
times, is evident.
So, at least from a conceptual point of view, one should regard this 
representation as  the fundamental one.  In principle,
one could even work in this representation to compute quantities such
as the spin correlation $C(t)$\/ in (\ref{Cqftdef}). However, 
since the values of matrix elements are not changed by unitary
transformations---in particular,
\be
\sqvb \, \wh{\Xi}_V(t)\, \sqvk =
\lla \psi_0 | \, \wh{\Xi}(t) \,| \psi_0 \rra
\ee
---it is  not necessary to do so.  The usual representation of
Heisenberg-picture quantum field theory, with initial-condition information
contained in the state vector, may thus be regarded as an auxiliary representation to which
one transforms from the explicitly-local vacuum representation
for purposes of computational convenience.

\section{DISCUSSION}

We see that the same mechanism that brings
about correlations at a distance in Heisenberg-picture quantum mechanics in a local
manner also works in Heisenberg-picture quantum field theory, at least in the nonrelativistic fermionic case.  In the vacuum representation of Sec. \ref{vacrepsec}, or
in any other representation in which the state vector is independent of initial conditions,
all information regarding initial values of
physical quantities is encoded in the initial-time field operators. Information encoded in an operator at one
place at an earlier time is transfered to operators at other places at later
times in accordance with a local differential equation. (In a relativistic theory
information is only transferred to later-time operators within the future light
cone of the earlier-time operator.) Operator-valued wave packets corresponding to initially-separated particles may come into contact and exchange information. At any time,
at any location, the value of the field operator is a weighted sum of products
of initial-time field operators, as in (\ref{Vpertsol}), (\ref{VJpertsol})  (of
course higher-order terms will in general be present).  As one wave packet
passes by another it may acquire contributions to this weighted sum which were ``carried''
to the interaction region by another wave packet corresponding to another particle. 
(E.g., via the $O(\epsilon)$\/ change to 
$\widehat{\phi}_{V,[r]i}$\/ due to the interaction term
in eq. (\ref{Vpertsol}).)
At later times operators in this wave packet will retain these contributions, serving as 
labels
indicating that the encounter with the other wave packet took place. Depending on
the nature of the initial conditions and the interaction, distant field operators
at times subsequent to the  interaction may be entangled in such a way that 
the results of measurements
made upon them (when compared at still later times by means of some other 
causal interaction) are correlated to a degree in excess of that allowed
by Bell's theorem.\footnote{Entanglement and violation of Bell's theorem are
related  but not equivalent phenomena.\cite{MunroNemotoWhite01}}  
As discussed in Ref. 22, %\cite{Rubin01}, 
correlations in the
Everett interpretation are correlations of
information exchanged in a causal manner
between copies of measuring instruments and/or
the states of awareness of observers, so these 
excess correlations in no way imply the presence 
of nonlocality.

One may be inclined to ask: 
``Since, in this representation, information is 
transported according to the dictates of a local 
differential equation, what additional  role is 
played by the Everett interpretation in rendering 
the theory local? I know about Bell's 
theorem; but what precisely {\em is}\/ it that goes 
wrong to spoil  the locality which is so evident in 
the differential equations, if I take the view 
that only a single outcome occurs to every 
measurement? Through what back door in the formalism 
does nonlocality sneak in?''
 
     The information which is transported locally 
in the quantum operators determines the possible 
outcomes of measurements and their respective 
probabilities.  In the Everett interpretation these 
possible outcomes and probabilities, completely 
computable from the locally-transported information, 
completely characterize the results of measurements.
In interpretations in which only a {\em single}\/
outcome occurs in measurement-type interactions,
additional information must be {\em added}\/
to completely characterize the outcome of 
measurement-type interactions: namely, {\em which 
outcome actually occurred}.\/  This information,
of course, is {\em  not}\/ carried in the quantum 
operators.

Since the issue of locality or the lack thereof is generally
framed in the EPRB experiment in terms of particles which are distinguishable,
at least by virtue of their spatial locations, 
and since the goal of the present paper has been to understand how the mechanism which permits
locality in the first-quantized theory
operates in the field-theoretic formalism,
the focus above has been on  states consisting of particles of distinct species, which
can always in principle be 
distinguished.  This is not to say that the issue of
indistinguishability, although distinct from that of locality,
does not impact locality.\footnote{The connection between entanglement
and particle indistinguishability is examined in Refs. 34-48.}
%\cite{SchleimannLossMacDonald01}-\cite{PachosSolano02}}
Indeed, it is difficult to see 
how  the labeling mechanism would operate causally in the context of symmetrized
or antisymmetrized first-quantized Hilbert space. In field theory, as seen
above, it is the field operators which are labeled by transformations induced by
interactions, so there is no  problem in
this regard.

As for the label-proliferation problem, quantum field theory  provides  no
explanation, beyond that provided by first-quantized theory, for the manner in which this information is recorded.
The representation of 
the label information does seems  more natural in quantum field theory than in 
point-particle quantum mechanics.  In quantum mechanics, the operators pertaining
to each particle acquire
tensor-product factors acting in the state spaces of other particles
with which the particle in question  interacts.  In quantum field theory,
each operator acts in an infinite-dimensional space  (see, e.g., Ref. 49 
%\cite{BjorkenDrellII} 
for an explicit representation) and
changes the nature of its action in this space based on the nature of nearby
operators. But, be it quantum-mechanical or quantum-field-theoretic,
a single quantum operator is capable of carrying
an unlimited amount of information regarding past 
interactions.\footnote{Kent\cite{Kent02} presents a hidden-variable
model in which ``every hidden particle trajectory carries a record of
its past history---which particles it was originally entangled with, and which measurements were
carried out on it---arbitrarily far into the future.'' This is precisely
the  information which, in  quantum theory {\em sans}\/ hidden variables,
particles do indeed 
carry with them.  See also Ref. 51.} %\cite{GalvaoHardy02}.}

\section*{ACKNOWLEDGMENTS}

I would like to thank Jian-Bin Mao, Allen J. Tino and Lev Vaidman for helpful
discussions.  I am especially grateful to
Rainer Plaga for persistently insisting that I clarify the presentation.

\end{document}